\documentclass[a4paper,aps,prc,preprintnumbers,eqsecnum,twocolumn,nofootinbib,notitlepage]{revtex4-1}
\usepackage{graphicx}
\usepackage{bm}
\usepackage{amsmath}
\usepackage{amssymb}
\usepackage{amsfonts}
\usepackage[dvips]{color}
\newcommand{\half}{\frac{1}{2}}

\newcommand{\der}{\partial}

\newcommand{\Tr}{\mbox{\rm Tr}}
\newcommand{\dsl}{\partial\kern-0.55em\raise 0.14ex\hbox{/}}

\newcommand{\bfk}{\bm{k}}

\newcommand{\bfp}{\bm{p}}

\newcommand{\bfx}{\bm{x}}
\newcommand{\bfy}{\bm{y}}

\newcommand{\bfA}{\bm{A}}

\newcommand{\bfD}{\bm{D}}

\newcommand{\bfP}{\bm{P}}

\newcommand{\bfsigma}{\bm{\sigma}}

\newcommand{\Slash}[1]{{\ooalign{\hfil/\hfil\crcr$#1$}}} 
\begin{document}

\preprint{KYUSHU-HET-129}

\title{Pions are neither perturbative nor nonperturbative: \\
Wilsonian renormalization group analysis of nuclear effective field
theory \\
including pions}

\author{Koji Harada}
\email{harada@phys.kyushu-u.ac.jp}
\affiliation{Department of Physics, Kyushu University\\
Fukuoka 810-8560 Japan}
\author{Hirofumi Kubo}
\email{kubo@higgs.phys.kyushu-u.ac.jp}
\affiliation{Synchrotron Light Application Center, Saga University \\
1 Honjo, Saga 840-8502, Japan}
\author{Yuki Yamamoto}
\email{yamamoto@koeki-u.ac.jp}
\affiliation{
Department of Community Service and Science,
Tohoku University of Community Services and Science\\
Iimoriyama 3-5-1, Sakata 998-8580 Japan}

\date{\today}

\begin{abstract}
 Pionful nuclear effective field theory (NEFT) in the two-nucleon sector
 is examined from the Wilsonian renormalization group point of view. The
 pion exchange is cut off at the floating cutoff scale, $\Lambda$, with
 the short-distance part being represented as contact interactions in
 accordance with the general principle of renormalization. We derive the
 nonperturbative renormalization group equations in the leading order of
 the nonrelativistic approximation in the operator space including up to
 $\mathcal{O}(p^2)$ and find the nontrivial fixed points in the $^1S_0$
 and $^3S_1$--$^3D_1$ channels which are identified with those in the
 pionless NEFT. The scaling dimensions, which determine the power
 counting, of the contact interactions at the nontrivial fixed points
 are also identified with those in the pionless NEFT. We emphasize the
 importance of the separation of the pion exchange into the
 short-distance and the long-distance parts, since a part of the former
 is nonperturbative while the latter is perturbative.
\end{abstract}


\maketitle

\newcommand{\comment}[1]{{\color{red}\tiny #1}}


\section{Introduction}


Nuclear effective field theory (NEFT) is the low-energy effective field
theory~\cite{Weinberg:1978kz} of nucleons based on symmetries of QCD,
and is expected to give a model-independent description of nuclear
phenomena at low energies.  Since the seminal papers of
Weinberg~\cite{Weinberg:1990rz, Weinberg:1991um, Weinberg:1992yk}, a lot
of applications have been done with success for more than two decades.
See Refs~\cite{Beane:2000fx, Bedaque:2002mn, Epelbaum:2005pn,
Machleidt:2007ms, Epelbaum:2008ga} for reviews. It still has a wide
range of phenomena to be explored. 


In spite of these successes, however, there are unsettled issues at the
fundamental level even in the simplest two-nucleon sector; whether the
pions should be treated perturbatively or not and how to treat the
contact interactions.  Because the understanding of the two-nucleon
scattering is essential also for other phenomena, this issue is of
central importance in NEFT.


The original power counting due to Weinberg~\cite{Weinberg:1990rz,
Weinberg:1991um, Weinberg:1992yk} for the ``effective potential'' is
nothing but the naive dimensional analysis. The ``effective potential''
is plugged into the Lippmann-Schwinger equation and the scattering
amplitude is calculated, so that the pion exchange is iterated infinite
times. In this Weinberg scheme the pions are thus treated
nonperturbatively. Kaplan, Savage, and Wise~\cite{Kaplan:1996xu} pointed
out that the Weinberg power counting scheme is inconsistent, in the
sense that the higher order counterterms are required to cancel the
divergences at the given order. They (and, independently, van Kolck)
proposed an alternative power counting scheme in which only the
non-derivative contact interaction, the $C_0$-term, is treated
nonperturbatively so that the pions are treated
perturbatively~\cite{Kaplan:1998tg, Kaplan:1998we,
vanKolck:1998bw}. This scheme (widely known as KSW scheme) is free from
the inconsistency, but Fleming, Mehen, and Stewart~\cite{Fleming:1999ee}
showed that the KSW scheme does not lead to the converging results in
the channels in which the singular tensor part of the pion exchange
contributes.  Beane, Bedaque, Savage, and van Kolck~\cite{Beane:2001bc}
proposed a remedy, known as the hybrid approach, in which the amplitudes
are expanded around the chiral limit so that the $1/r^3$ part (which
survives in the limit) is treated nonperturbatively.


The treatment of the tensor part of the one-pion exchange (OPE) is the
source of controversy. See also Refs.~\cite{Nogga:2005hy,
Epelbaum:2006pt, Hammer:2006qj, Birse:2005um, Gegelia:2004pz,
Long:2007vp} for more details.


Recently, Beane, Kaplan, and Vuorinen (BKV)~\cite{Beane:2008bt}
considered a Pauli-Villars type regularization for the OPE in order to
separate the short-distance part of the tensor interaction from the
long-distance part. The short-distance part is represented by contact
interactions. They employed the power divergence subtraction (PDS)
renormalization and obtained the convergent results.


Note that not all the $1/r^3$ part of the OPE is singular. It is the
\textit{short-distance part} that is singular and spoils the convergence
of the amplitude as found in Ref~\cite{Fleming:1999ee}. It means that
all the $1/r^3$ part of the OPE does not need to be iterated.


The idea of separating the singular short-distance part of the tensor
interaction from the long-distance part and representing the former by
contact interactions (BKV prescription) is essentially the Wilsonian
renormalization group (RG) idea of effective
interactions~\cite{Wilson:1973jj, Wilson:1974mb}. The separation scale
in Ref.~\cite{Beane:2008bt} may be viewed as an analog of the floating
cutoff in the Wilsonian RG.


The Wilsonian RG is a useful tool to investigate the effects of the
short-distance physics on the long-distance physics. When the quantum
fluctuations with momenta higher than the floating cutoff $\Lambda$ are
integrated out, their effects are simulated by a series of local
operators, which serve as low-momentum effective interactions. The
coupling constants thus depend on the cutoff $\Lambda$, while the
physical quantities such as scattering amplitudes do not.



The Wilsonian RG has been applied to NEFT in order to examine the power
counting issue~\cite{Birse:1998dk, Barford:2002je, Birse:2005um,
Birse:2008wt, Harada:2005tw, Harada:2006cw, Harada:2007ua}. (See also
Ref.~\cite{Nakamura:2004ek, Nakamura:2006hc, Kvinikhidze:2007eu} for
other use of the Wilsonian RG in NEFT.) The existence of the nontrivial
fixed points of the RG equations (RGEs) in the S-waves explains
unnaturally large scattering lengths. The power counting can be
determined by the scaling dimensions of operators around the fixed
points. In particular, it is relevant operators that should be resummed
to all orders.


Birse~\cite{Birse:2005um} examines the pionful NEFT by using the
so-called ``distorted-wave RG~\cite{Barford:2002je}'' and claims that
the scaling dimensions are shifted from those of the pionless theory
due to the singularity of the tensor force in the spin-triplet channel.
Also, his results imply that the effects of pions do not decouple even
at very low momenta.


It sounds strange however that pions do not decouple at the momenta
where the pionless NEFT is valid. The effects should be eventually
represented by contact interactions at very low momenta.  One should
expect that the transition from the pionful NEFT to the pionless NEFT is
smooth. A formulation of the Wilsonian RG which permits the smooth
transition is desired.


In this paper, we perform the Wilsonian RG analysis of the pionful NEFT
in the two-nucleon sector, and examine the effects of pions. In the
Wilsonian RG approach, there is a single scale $\Lambda$ which separates
the long-distance physics from the short-distance physics so that one
does not need to implement an additional regularization for the
short-distance part of the OPE, as does in the BKV prescription.


The key idea here is the separation
of the short-distance part from the long-distance part. NEFT, as a
low-energy effective theory, has a finite range of applicability,
specified by the physical cutoff $\Lambda_0$. It is a general principle
of EFT that the physics with the momentum scale larger than $\Lambda_0$
is represented as local operators. The short-distance part of the OPE
(S-OPE) should also be represented as local operators (contact
interactions). (In the Wilsonian RG analysis, the floating cutoff
$\Lambda$ plays the role of $\Lambda_0$, after integrating the modes
with momenta between $\Lambda$ and $\Lambda_0$.)


Note that, although the contact interactions are attributed to the
effects of the heavy particles in the usual EFT lore, this is not
precise. They arise also from the short-distance physics of light
particles, such as pion exchanges.  Note also that the representation of
short-distance physics by local interactions is a general treatment and
has nothing to do with how singular the S-OPE is.


We find nontrivial fixed points in the $^1S_0$ and $^3S_1$--$^3D_1$
channels, which are responsible for the large scattering
lengths. Importantly these fixed points are identified with those found
in the Wilsonian RG analysis in the pionless NEFT. Thus the scaling
dimensions at the fixed points are also the same. There is one relevant
operator in each channel. That is, the pion interactions do not alter
the scaling dimensions.


We emphasize that the question of whether the pions are perturbative or
nonperturbative is not well posed.  Since the S-OPE is represented as
contact interactions in the Wilsonian analysis, only the long-distance
part of the OPE (L-OPE) is the proper interactions due to OPE. On the
other hand, the S-OPE cannot be distinguished from other contributions,
such as heavier meson ($\rho$, $\omega$, etc.)  exchanges. (Because of
the cutoff, we do not have enough resolution.)  The RGEs tell us that,
at the nontrivial fixed point, the L-OPE should be treated
perturbatively, while there is a relevant operator (a part of which is
the S-OPE) that should be resummed to all orders, i.e.,
nonperturbatively.


The existence of the relevant operator at the nontrivial fixed point
sharpens the distinction between the S-OPE and the L-OPE. If there were
no relevant operator, the separation would not make much difference.


Our Wilsonian RG permits the smooth transition from the pionful NEFT to
the pionless NEFT. The nontrivial fixed points do not change nor the
scaling dimensions. The L-OPE transmutes into contact interactions which
represent the S-OPE as the cutoff is lowered. At the value of $\Lambda$
lower than the pion mass, the most of the L-OPE has been changed into
contact interactions, thus the system is well described by the pionless
NEFT.


The structure of the paper is the following. In Sec~\ref{sec:WRG} we
recapitulate the main points of the previous papers in order to
introduce the notations and the main concepts in our analysis. In
Sec.~\ref{sec:IR} we discuss the nonrelativistic approximation. Starting
with the nonrelativistic nucleons and the relativistic pions, we
estimate the order of magnitude of a various kind of diagrams
contributing to the RGEs and determine the leading terms. In
Sec.~\ref{sec:pionful} we present the RGEs and examine the structure of
the flows in the $^1S_0$ and $^3S_1$--$^3D_1$ channels. The nontrivial
fixed points are found to be identified with those of the pionless
theory.  Sec~\ref{sec:summary} is devoted to the summary and the
comments on the related works. In Appendix~\ref{sec:lowlambda} the RGEs
for the case of $\Lambda < m_{\pi}$ are presented and compared with
the pionless case. In Appendix~\ref{sec:qed} we discuss the similarity
and the difference between the pionful NEFT and QED.

\section{Wilsonian RG analysis of Pionless NEFT in two-nucleon sector}
\label{sec:WRG}

In the previous papers~\cite{Harada:2005tw, Harada:2006cw,
Harada:2007ua}, we have explained the basic concepts of the Wilsonian RG
and its relevance to the power counting in the NEFT. Here, we briefly
recapitulate it, give some remarks, and introduce the notations used in
later sections.

\subsection{What is the use of the Wilsonian RG in NEFT?}


The most basic idea behind the power counting is the order of magnitude
estimate based on dimensional analysis. In an EFT with the physical
cutoff $\Lambda_0$, the dimensional analysis is usually based on this
scale. At the classical level, the (canonical) mass dimension of an
operator is determined with respect to the kinetic term. For example, a
Dirac field $\psi$ with a kinetic term $\mathcal{L}_{kin} = \bar{\psi}
i\dsl\psi$ has mass dimension three halves, $[\psi]=3/2$. The dimensions
of other operators are determined accordingly. The operator
$(\bar{\psi}\psi)^2$ has dimension six, so that it enters in the
Lagrangian as $(c/\Lambda_0^2)(\bar{\psi}\psi)^2$, where $c$ is a
dimensionless constant. The coupling constant $c/\Lambda_0^2$ associated
with the operator $(\bar{\psi}\psi)^2$ has dimension $-2$, which counts
the power of $\Lambda_0$.



Quantum fluctuations may, in general, change the classical dimensional
analysis. The quantum counterpart of the (canonical) dimension is
called the \textit{scaling dimension}, which can be obtained by the RG
analysis. Wilsonian RG is a nonperturbative tool to
handle the quantum fluctuations.


An operator whose coupling has a negative (scaling) dimension is called
\textit{irrelevant} because it becomes less important at lower
energies. An operator whose coupling has a positive (scaling) dimension
is called \textit{relevant} because it becomes more important at lower
energies. An operator whose coupling is dimensionless is called
\textit{marginal}. The scaling dimension of an operator is the measure
of how important it is. It is therefore natural to consider the power
counting on the basis of the scaling dimensions.


In the S-wave scattering of two nucleons, the scattering lengths are
known to be much larger than the ``natural'' size, $1/\Lambda_0$. (In
the case of the pionless NEFT, the physical cutoff is of order of the
pion mass, $\Lambda_0\sim \mathcal{O}(m_{\pi})$.) From the RG point of
view, the ``fine-tuning'' is related to the existence of a nontrivial
fixed point (and a critical surface) of the RG flow.


Around the nontrivial fixed points the scaling dimensions are
drastically different from the canonical dimensions.  It has been
shown~\cite{Birse:1998dk, Harada:2006cw, Harada:2007ua} that the
coupling which corresponds to the scattering length becomes relevant,
although it is irrelevant at the classical level.


There are values of coupling constants with which the scattering length
is infinite. This set of coupling constants forms a critical surface. It
separates the weak-coupling and the strong-coupling phases. For the
two-nucleon system, the spin-singlet channel is considered to be in the
weak-coupling phase, while the spin-triplet channel is considered to be
in the strong-coupling phase, because of the (non)existence of a
bound state in these channels.


To summarize: the two-nucleon system with large S-wave scattering
lengths is governed by the existence of nontrivial fixed points, and the
Wilsonian RG is a systematic tool to study the scaling dimensions on
which the power counting should be based, around the nontrivial fixed
points.

\subsection{Remarks on Wilsonian RGEs with Galilean invariance}


There are several formulations for the Wilsonian RG
analysis~\cite{Wegner:1972ih, Polchinski:1983gv, Nicoll:1977hi,
Bonini:1992vh, Wetterich:1992yh, Morris:1993qb}, which are however
essentailly equivalent. The most popular one is the functional RG
method. (See Refs.~\cite{Berges:2000ew, Pawlowski:2005xe} for reviews.)
In this formulation, a cutoff function is introduced for each propagator
to suppress the low-frequency fluctuations. The effective averaged
action $\Gamma_{\Lambda}[\Phi]$, which interpolates the bare classical
action $S[\Phi]$ ($\Lambda=\Lambda_0$) and the effective action
$\Gamma[\Phi]$ ($\Lambda=0$), depends on the floating cutoff scale,
$\Lambda$, as
\begin{equation}
 \frac{d\Gamma_{\Lambda}}{d\Lambda} = \frac{1}{2}
  \Tr
  \left[
   \frac{dR_\Lambda}{d\Lambda}\left(\Gamma_{(2)}+R_\Lambda\right)^{-1}
  \right],
\end{equation}
where $\Phi$ is the classical field, $\Gamma_{(2)}$ stands for the
second derivative of the averaged action $\Gamma_\Lambda$ with respect
to $\Phi$, and $R_{\Lambda}$ is the cutoff function which suppresses the
fluctuations with $p \alt \Lambda$. Note that, although it is an
``one-loop'' equation, it contains all the nonperturbative information.


A straightforward application of this formulation to a nonrelativistic
system however encounters difficulties.  In the usual formulation for a
relativistic system, one considers the theory in Euclidean space. The
cutoff is imposed on the magnitude of four-momentum of the
propagator. On the other hand, in a nonrelativistic system, space and
time should be treated differently, and thus the Euclidean formulation
cannot be used. One might rather like to consider a cutoff imposed on
the three-momentum in the propagator. But such a cutoff necessarily
breaks Galilean invariance of the nonrelativistic system.  There is no
obvious way to impose a Galilean invariant cutoff at the averaged action
level. This is a general feature independent of the choice of the cutoff
function.  Furthermore, if the cutoff function is not smooth enough,
non-analytic terms in momenta arise\footnote{If one considers only a few
leading order terms in the derivative expansion, typically in the local
potential approximation, non-smooth cutoff does not cause the
problem. It is the reason why the problem of non-analyticity is not
revealed in many applications.}. See Ref.~\cite{Birse:2008wt} for an
example in a similar context.


In a system of \textit{two} nonrelativistic particles, there is a simple
and physically sensible way out of this problem: it is to impose a
cutoff on the \textit{relative} three-momentum of the two particles,
which is Galilean invariant.  In addition, the results are very
insensitive to the choice of the cutoff function. See Appendices of
Refs~\cite{Harada:2006cw, Harada:2007ua}. In particular, the results
with a sharp cutoff are the same as those with a smooth one. It is a
technical advantage that a sharp cutoff can be used because it
simplifies the calculations considerably.

\subsection{Pionless NEFT up to $\mathcal{O}(p^2)$ in the $^1S_0$ and
  $^3S_1$--$^3D_1$ channels}
\label{sec:pionless}


In Ref.~\cite{Harada:2006cw}, we consider the pionless NEFT without
isospin breaking. The relevant degrees of freedom are nonrelativistic
nucleons, which interact with themselves only through contact
interactions. In the two-nucleon sector, they are four-nucleon operators
with an arbitrary number of derivatives. An operator with more
derivatives has higher canonical dimensions than the ones with less
derivatives. 


Since there are infinitely many operators involved in the flow equation,
one needs to introduce a truncation of the space of operators in the
averaged action in order to solve it. We retain only the operators with
derivatives up to a certain order. We simply count the number of spatial
derivatives ($\nabla \sim p$) and a time derivative is counted as two
spatial derivatives ($\der_t \sim p^2$).  We consider the following
ansatz for the averaged action up to $\mathcal{O}(p^2)$,
\begin{widetext}
\begin{eqnarray}
\Gamma^{(\Slash{\pi})}_{\Lambda}&=&\int d^4x
 \bigg[
 N^{\dagger}\left(i\partial_t+\frac{{\nabla}^2}{2M}\right)N
 \nonumber \\
 &&\left\{
  \begin{array}{lcl}
   -C^{(S)}_0\mathcal{O}^{(S)}_0
    +C^{(S)}_2\mathcal{O}^{(S)}_2
    +2B^{(S)}\mathcal{O}^{(SB)}_2\bigg],
    &\quad& \mbox{(${}^1S_0$\ channel)}\\
   -C^{(T)}_0\mathcal{O}^{(T)}_0
    +C^{(T)}_2\mathcal{O}^{(T)}_2
    +2B^{(T)}\mathcal{O}^{(TB)}_2
    +C^{(SD)}_2\mathcal{O}^{(SD)}_2\bigg],
    &\quad& \mbox{(${}^3S_1$--${}^3D_1$\ channel)}\\
  \end{array}
\right.
\label{truncation}
\end{eqnarray}
where the operators in the ${}^1S_0$ are given by
\begin{subequations}
\begin{eqnarray}
 \mathcal{O}^{(S)}_0&=&
  \left(
   N^TP_a^{(S)}N
  \right)^{\dagger}
  \left(
   N^{T}P_a^{(S)}N
  \right), \\
 \mathcal{O}^{(S)}_2&=&
 \left[
  \left(
   N^TP_a^{(S)}N
  \right)^{\dagger}
  \left(
   N^{T}P_a^{(S)}\overleftrightarrow{\nabla}^2N
  \right)+ h.c.
 \right], \\
\mathcal{O}^{(SB)}_2&=&
\left[
 \left\{
  N^TP_a^{(S)}\left(i\partial_t+\frac{\nabla^2}{2M}\right)N
 \right\}^{\dagger}
 \left(
  N^TP_a^{(S)}N
 \right)+h.c.
\right], 
\end{eqnarray}
\end{subequations}
and in the ${}^3S_1$--${}^3D_1$ channel, 
\begin{subequations}
\begin{eqnarray}
 \mathcal{O}^{(T)}_0&=&
  \left(
   N^TP_i^{(T)}N
  \right)^{\dagger}
  \left(
   N^{T}P_i^{(T)}N
  \right),\\
 \mathcal{O}^{(T)}_2&=&
  \left[
   \left(
    N^TP_i^{(T)}N
   \right)^{\dagger}
   \left(
    N^{T}P_i^{(T)}\overleftrightarrow{\nabla}^2N
   \right)+ h.c.
  \right], \\
 \mathcal{O}^{(SD)}_2&=&
  \left[
   \left(
    N^TP_i^{(T)}N
   \right)^{\dagger}
   \left\{
    N^{T}
    \left(
     \overleftrightarrow{\nabla}_i\overleftrightarrow{\nabla}_j
     -\frac{1}{3}\delta_{ij}\overleftrightarrow{\nabla}^2
     \right)
    P^{(T)}_{j}N
   \right\}+ h.c.
   \right], \\
 \mathcal{O}^{(TB)}_2&=&
  \left[
   \left\{
    N^TP_i^{(T)}\left(i\partial_t+\frac{\nabla^2}{2M}\right)N
   \right\}^{\dagger}
   \left(
    N^TP_i^{(T)}N\right)+h.c.
  \right],\\
\end{eqnarray}
\end{subequations}
\end{widetext}
where we have introduced the notation $\overleftrightarrow{\nabla}^2
\equiv \overleftarrow{\nabla^2} + \overrightarrow{\nabla^2} -
2\overleftarrow{\nabla}\cdot\overrightarrow{\nabla}$ and the projection
operators\cite{Fleming:1999ee},
\begin{equation}
 P_a^{(S)}\equiv\frac{1}{\sqrt{8}}\sigma^2\tau^2\tau^a,
  \qquad
 P_k^{(T)}\equiv\frac{1}{\sqrt{8}}\sigma^2\sigma^k\tau^2,
\end{equation}
for the ${}^1S_0$ (spin singlet) channel and the ${}^3S_1$ (spin
triplet) channel respectively. The nucleon field $N(x)$ with mass $M$ is
an isospin doublet nonrelativistic two-component spinor. Pauli matrices
$\sigma^i$ and $\tau^a$ act on spin indices and isospin indices
respectively. The two channels are completely decoupled, and thus we can
consider each channel separately.


Note that the possible forms of the operators are restricted by Galilean
invariance. Note also that we have included ``redundant operators,''
$\mathcal{O}^{(SB)}_2$ and $\mathcal{O}^{(TB)}_2$, because they are
necessary in a consistent Wilsonian RG analysis~\cite{Harada:2005tw}.


We introduce the following dimensionless coupling constants,
\begin{equation}
x\equiv \frac{M\Lambda}{2\pi^2} C_0^{(S)},\quad
y\equiv \frac{M\Lambda^3}{2\pi^2} 4C_2^{(S)},\quad
z\equiv \frac{\Lambda^3}{2\pi^2} B^{(S)},
\label{dimensionlesssinglet}
\end{equation}
for the spin-singlet channel and 
\begin{eqnarray}
x'&\equiv& \frac{M\Lambda}{2\pi^2} C_0^{(T)}, \quad
y'\equiv \frac{M\Lambda^3}{2\pi^2}4 C_2^{(T)}, \quad 
z'\equiv \frac{\Lambda^3}{2\pi^2} B^{(T)}, \nonumber \\
w'&\equiv& \frac{M\Lambda^3}{2\pi^2}\frac{4}{3}C_2^{(SD)},
\label{dimensionlesstriplet}
\end{eqnarray}
for the spin-triplet channel. With the sharp cutoff on the relative
momenta, the flow equations that determine the dependence on
$t=\ln(\Lambda_0/\Lambda)$ of the coupling constants can be written
as~\cite{Harada:2007ua}
\begin{equation}
 \frac{dx_C}{dt} + d_C x_C
  =\left.\sum_{A,B} x_A x_B
  \frac{M\Lambda}{2\pi^2}
  \frac{F_A(p_i,\Lambda)F_B(\Lambda,p_f)}
  {1-\tilde{A}(p_i)}\right|_C,
  \label{sharpRGE}
\end{equation}
where $x_C$ stands for one of the dimensionless coupling constants, and
$d_C$ is the power of $\Lambda$ in the definition of the dimensionless
coupling constant. In the following, we call $- d_C$ the canonical
dimension~\footnote{This definition of the canonical dimension reflects
the nonrelativistic scaling property. See Ref.~\cite{Luke:1996hj}.} of
the coupling. $\tilde{A}(P)$ is defined as
\begin{equation}
 A(P)=P^0-\frac{\bfP^2}{4M},\ 
  \tilde{A}(P)=M A(P)/\Lambda^2, 
\end{equation}
where $P=(P^0, \bfP)$ is the total momentum of the system. $F_A(p_i,
p_f)$ is the momentum-dependent factor associated with the coupling
$x_A$:
\begin{align}
 F_x&=\frac{-2\pi^2}{M\Lambda}, \quad
 F_y=\frac{-2\pi^2}{4M\Lambda^3}\left(r_{12}+r_{34}\right), 
 \nonumber \\
 F_z&=\frac{2\pi^2}{\Lambda^3}\sum_{i=1}^4 S_i, 
\end{align}
for the spin-singlet channel, and
\begin{align}
 F_{x'}&=\frac{-2\pi^2}{M\Lambda}, \quad
 F_{y'}=\frac{-2\pi^2}{4M\Lambda^3}\left(r_{12}+r_{34}\right), 
 \nonumber \\
 F_{z'}&=\frac{2\pi^2}{\Lambda^3}\sum_{i=1}^4 S_i, 
 \nonumber \\
 {F}^{ij}_{w'}&=\frac{3}{4}\left(\frac{-2\pi^2}{M\Lambda^3}\right)
 \left[
 p_{12}^i\; p_{12}^j + p_{34}^i\; p_{34}^j 
 -\frac{1}{3}\delta^{ij} (r_{12}+r_{34})
 \right],
\end{align}
for the spin-triplet channel,
with
\begin{equation}
 S_i=p_i^0-\frac{\bfp_i^2}{2M}, \
  \bfp_{ij}=\bfp_{i}-\bfp_{j}, \ 
  r_{ij}=\left(\bfp_i-\bfp_j\right)^2.
\end{equation}
$\bfp_1$ and $\bfp_2$ are the incoming momenta of the nucleons to the
vertex, and $\bfp_3$ and $\bfp_4$ are the outgoing momenta from the
vertex.  The notation $|_C$ stands for the operation of taking the
coefficient of $F_C(p_i,p_f)$ in the sum.


By using the formula Eq.~(\ref{sharpRGE}) the RGEs are obtained. The
explicit expressions are not presented here, but can be read easily from
the RGEs (Eqs.~(\ref{RGEsinglet}) and Eqs.~(\ref{RGEtriplet})) for the
pionful theory discussed in Sec.~\ref{sec:pionful}.


There is a nontrivial fixed point in each channel, which is relevant to
the physical two-nucleon system:
\begin{equation}
 \left(x^{\star},y^{\star},z^{\star}\right)=
  \left(-1,-\frac{1}{2},\frac{1}{2}\right),
  \label{ntfpsinglet}
\end{equation}
in the spin-singlet channel, and 
\begin{equation}
\left({x'}^{\star},{y'}^{\star},{z'}^{\star},{w'}^{\star}\right)=
\left(-1,-\frac{1}{2},\frac{1}{2},0\right),
\label{ntfptriplet}
\end{equation}
in the spin-triplet channel. At the nontrivial fixed point, the
operators get large anomalous dimensions and there is one operator that
becomes \textit{relevant} with the scaling dimension of the coupling
constant being one in each channel.

\section{Nonrelativistic approximation, IR enhancement and the leading
 order in $\Lambda/M$}
\label{sec:IR}


In this section, we consider the inclusion of pions as dynamical degrees
of freedom. It extends the range of applicability of NEFT to higher
momenta beyond the pion mass scale. The contact interactions in the
pionless theory resolve into the effects by pion propagation and the
rest. The physical cutoff $\Lambda_0$ is now larger than $m_{\pi}$, and
we suppose that $\Lambda_0$ is of order $400$ MeV.

\subsection{Chiral symmetry and nonrelativistic nucleons}


The most important feature of the pionful theory is chiral symmetry,
$SU(2)_{L}\times SU(2)_{R}$ spontaneously broken to $SU(2)_{V}$. It is
convenient to introduce the field $\Sigma$, which transforms linearly as
\begin{equation}
 \Sigma(x) \to L\Sigma(x)R^\dagger,
\end{equation}
where $L$ and $R$ are the elements of $SU(2)_{L}$ and $SU(2)_{R}$
respectively. The pion field $\pi^a(x)$ may be defined through
\begin{equation}
 \Sigma(x)= \exp(i\pi^a(x)\tau^a/f),
\end{equation}
where $f$ is the pion decay constant in the chiral limit.
The nucleon field transforms as
\begin{equation}
 N(x) \to U(x) N(x),
\end{equation}
where $U(x)$ is a function of $L$, $R$, and $\Sigma(x)$ and defined
through
\begin{equation}
 \xi(x)\to L\xi(x)U(x)^\dagger=U(x)\xi(x)R^\dagger,
  \label{xi}
\end{equation}
where $\xi^2(x)=\Sigma(x)$, i.e., $\xi(x)=\exp(i\pi^a(x)\tau^a/2f)$.


The chiral invariant Lagrangian for the nonrelativistic nucleon
interacting with pions is given as
\begin{eqnarray}
 \mathcal{L}_{NR} &=& 
  N^\dagger 
  \left[
   iD_0 + \frac{\left(\bfsigma\cdot \bfD\right)^2}{2M}
  \right]N
  +g_A N^\dagger \bfsigma \cdot \bfA N
  \nonumber \\
 && -C_0^{(c)} \mathcal{O}_{0}^{(c)}+C_2^{(c)}\mathcal{O}_{2}^{(c)} 
  +2B^{(c)} \mathcal{O}_{2}^{(cB)}
  \nonumber \\
 &&-D_2^{(c)} 
  \frac{m_{\pi}^2}{2} \Tr(\Sigma + \Sigma^\dagger) \mathcal{O}_{0}^{(c)}
  \nonumber \\
 &&+ \cdots,
  \label{lagNR}
\end{eqnarray}
where the chiral covariant derivative $D_{\mu}$ is defined as
\begin{equation}
 D_\mu N =(\der_\mu + V_\mu) N, 
\end{equation}
and $V_{\mu}$ and $A_{\mu}$ are defined as
\begin{eqnarray}
 V_\mu&\equiv&\half
  \left(\xi^\dagger \der_\mu \xi+\xi\der_\mu \xi^\dagger \right), \\
 A_\mu&\equiv&\frac{i}{2}
  \left(\xi^\dagger \der_\mu \xi-\xi\der_\mu \xi^\dagger \right).
\end{eqnarray}
The superscript $(c)$ denotes the channel to be specified, and the
summation is also implied, i.e., the term
$C_2^{(c)}\mathcal{O}_{2}^{(c)}$ for the spin-triplet channel contains
$C_2^{(T)}\mathcal{O}_{2}^{(T)}$ and $C_2^{(SD)}\mathcal{O}_{2}^{(SD)}$.
Note that the derivatives in the four-nucleon operators should be
replaced by the covariant derivatives, though we do not explicitly show
them here.


The ellipsis in Eq.~(\ref{lagNR}) denotes other terms, e.g., the higher
order operators, including six-nucleon operators, four-nucleon operators
with more than two derivatives, etc. It also contains the counterterms
of the form $N^\dagger N$ and $N^\dagger \nabla^2 N$, which play an
important role in the renormalization of the nucleon mass. Chiral
symmetry does not prevent the appearance of the operators that contain
more than one $\bfA$, such as $N^\dagger \bfA^{2} N$.  But it is not
generated by nonrelativistic reduction from the simple (chiral
invariant) Dirac action.  It implies that the operators have smaller
coefficients than $1/M$. Furthermore, we will see that it either does
not contribute to the RGEs for the four-nucleon operators, or gives only
suppressed contributions to the order we are working.


The Lagrangian for the pions is given by
\begin{equation}
 \mathcal{L}_{\pi}=\frac{f^2}{4}
  \left[
   \Tr\left(\der_\mu \Sigma^\dagger \der^\mu \Sigma\right)
   +m^2_{\pi}\Tr\left(\Sigma^\dagger + \Sigma \right)
\right] + \cdots.
\label{lpi}
\end{equation}

\subsection{Order of magnitude estimation of the contributions to the 
Wilsonian RGEs}
\label{sec:order}


In order to perform the Wilsonian RG analysis, one usually needs to use
a cutoff function that preserves all the symmetries of the
theory. Unfortunately, there does not seem to exist a manifestly chiral
invariant cutoff function which controls all the fluctuations, because
of the nonlinearity of the transformation~\cite{HHKNY}. Furthermore, it
is known that perturbation theory generates apparently noninvariant
terms (ANTs) even with the lattice and the dimensional regularization,
which preserve chiral symmetry~\cite{Harada:2009nb,
Harada:2009uua}. ANTs are also expected to appear in the Wilsonian RG
analysis, but it is not obvious how to treat them.


In addition, the inclusion of pions makes the notion of relative
momentum obscure. Furthermore, because pions are relativistic, Galilean
invariance does not make good sense as a constraint.


Fortunately, however, it turns out that these problems do not interfere
with the leading order calculations in the nonrelativistic
approximation. The argument is based on the order of magnitude estimate
of the contribution of each diagram to the Wilsonian RGEs.


We are going to obtain the RGEs for the two-nucleon sector in the
next-to-leading order in momentum expansion (to $\mathcal{O}(p^2)$) of
the averaged action in the nonrelativistic formulation. Let us first
remind the following things:
\begin{itemize}
 \item The contributions to the Wilsonian RGEs only come from one-loop
       diagrams.
 \item Because of the nonrelativistic feature, diagrams with
       anti-nucleon lines are absent. Thus, the diagrams are divided
       into sectors, each of them is specified by the nucleon
       number. The $n$-nucleon sector has $n$ nucleons at each time
       slice.
 \item The contributions to the two-nucleon sector do not come from
       $n$-nucleon operators with $n\ge 6$.
 \item Our primary concern is the renormalization of the four-nucleon
       operators with no pion emission and absorption. Chiral symmetry
       constrains the renormalization of the operators in which the pion
       fields appear through covariant derivatives.
 \item Supplementarily, one needs to consider the self-energy diagram of
       the nucleon and the diagrams for the nucleon-pion vertex that
       contribute to the RGEs for the four-nucleon operators.
\end{itemize}

Note also that there is no contributions to the pion self-energy to this
order.

\subsubsection{Four-nucleon operators}

\begin{figure}
 \includegraphics[width=0.9\linewidth,clip]{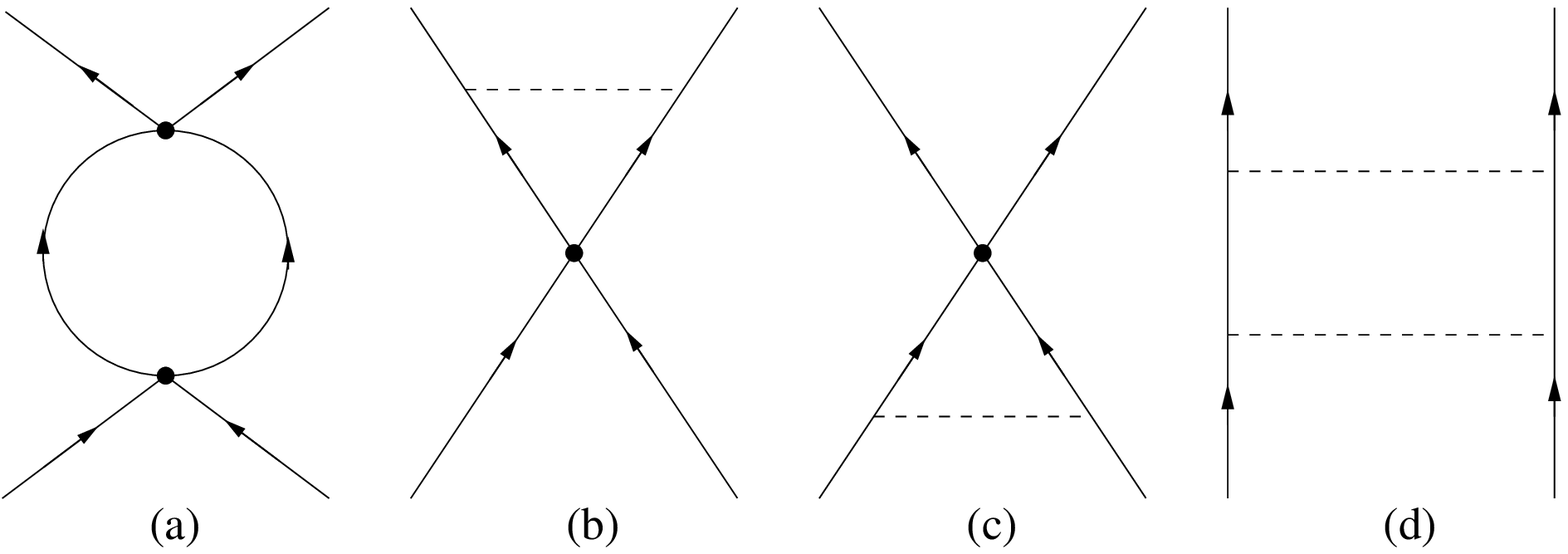}

 \vspace{0.5cm}

 \includegraphics[width=0.9\linewidth,clip]{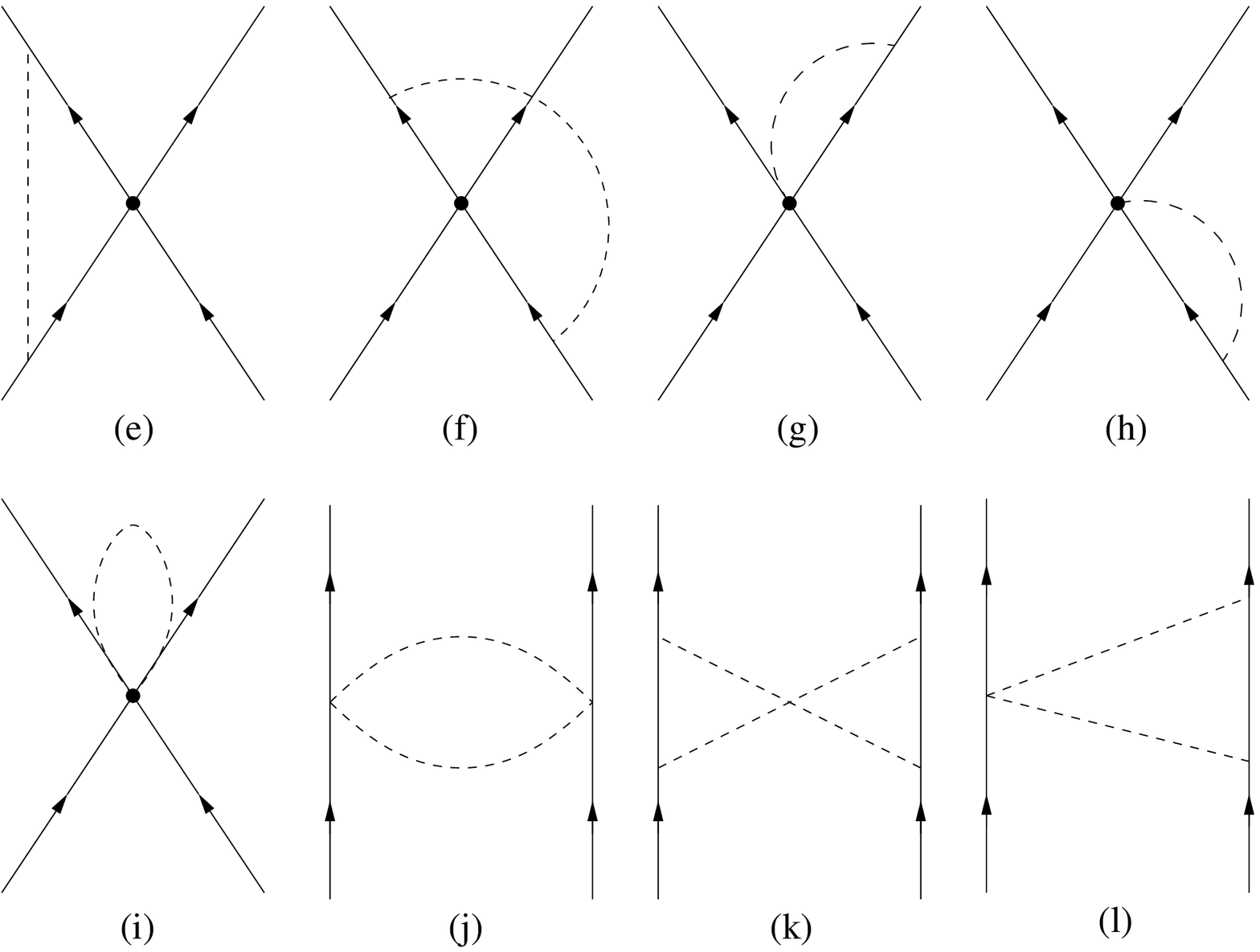}
 \caption{\label{fournucleon}Contributions to the four-nucleon
 operators. The four-nucleon vertices represent contact interactions
 collectively. The dotted lines represent the pion propagators. For the
 diagrams (e), (f), (g), (h), and (l), there are also mirrored diagrams
 with the left and the right interchanged.}
\end{figure}

According to the above-mentioned remarks, we need to consider the
diagrams given in Fig.~\ref{fournucleon} for the four-nucleon operators.

In order to obtain the RGEs, we need to evaluate the contributions from
the so-called ``shell-mode,'' the loop-contributions with the magnitude
of the loop (relative) three-momentum $k =|\bfk|$ is between
$\Lambda-d\Lambda$ and $\Lambda$. The argument is similar to the ``power
counting'' with the scale $Q$ found in the literature, but here, (i) we
do not consider the amplitude but the contribution to the RGEs, and (ii)
the magnitude of the momentum in the loop is actually $\Lambda$, while
in the ``$Q$-counting'' it is \textit{assumed} that the dominant
contributions come from the loop momentum of order $Q$.


To see how the dominant contributions to the RGEs emerge in certain
diagrams, let us consider the diagram (b) in Fig.~\ref{fournucleon} as
an illustrative example. At the moment, we assume that the floating
cutoff $\Lambda$ is larger than $m_{\pi}$ for simplicity.  The diagram
contains the loop integral
\begin{eqnarray}
 \int_{\mbox{\scriptsize shell}} 
  \frac{d^4k}{(2\pi)^4}
  &&
  \frac{i\bfk^2}{k^2-m^2_{\pi}+i\epsilon}
  \frac{i}{(p^0+k^0)-(\bfp+\bfk)^2/2M +i\epsilon}
  \nonumber \\
 &&\times
  \frac{i}{({p'}^0-k^0)-(\bfp'-\bfk)^2/2M +i\epsilon},
  \label{triangle1}
\end{eqnarray}
where $\int_{\scriptsize\mbox{shell}}$ denotes the shell mode integral
with the restriction on the magnitude of the relative three-momentum.
 There are four poles in the complex
$k^0$ plane, one nucleon and one pion poles in both the upper and the
lower half planes. One can evaluate the integral by the residues of the
poles in either half plane, and finds that the nucleon pole gives a
dominant contributions,
\begin{eqnarray}
 \int_{\mbox{\scriptsize shell}} \!
  \frac{d^3k}{(2\pi)^3}
  &&
  \frac{-\bfk^2}
  {\left[-p^0+(\bfp+\bfk)^2/2M\right]^2-\omega_{k}^2}
  \nonumber \\
 &&\times
  \frac{1}{(p^0\!+\!{p'}^0)-(\bfp\!+\!\bfk)^2/2M-(\bfp'\!-\!\bfk)^2/2M}
  \nonumber \\
 \sim \int_{\mbox{\scriptsize shell}} 
  \frac{d^3k}{(2\pi)^3}
  &&
  \frac{-(\bfk-(\bfp-\bfp')/2)^2}
  {\left[-p^0+(\bfk+\bfP/2)^2/2M\right]^2-\omega_{k-(p-p')/2}^2}
  \nonumber \\
 &&\times
  \frac{1}{E-\bfP^2/4M-\bfk^2/M}
\end{eqnarray}
where $\omega_{k}\equiv \sqrt{\bfk+m_{\pi}^2}$, and $E\equiv p^0+{p'}^0$
and $\bfP \equiv \bfp +\bfp'$ are the total energy and the total
three-momentum of the two nucleons, respectively. In going to the second
line, we have made a shift of the integration momentum so that $\bfk$ is
now the relative momentum. Since $M \gg |\bfk|=\Lambda \gg |\bfP|,\;
p^0$, and $E-\bfP^2/4M \ll \Lambda^2/M$, one may estimate the loop
integral as
\begin{equation}
 \sim \frac{-1}{2\pi^2}M d\Lambda.
\end{equation}
The pion pole gives $\sim \Lambda d\Lambda/4\pi^2$, which is smaller
than the dominant contribution by a factor of $\Lambda/M$.

On the other hand, the diagram (e) in Fig.~\ref{fournucleon}
contains
\begin{eqnarray}
 \int_{\mbox{\scriptsize shell}} 
  \frac{d^4k}{(2\pi)^4}
  &&
  \frac{i\bfk^2}{k^2-m^2_{\pi}+i\epsilon}
  \frac{i}{(p^0-k^0)-(\bfp-\bfk)^2/2M +i\epsilon}
  \nonumber \\
 &&\times
  \frac{i}{({p'}^0-k^0)-(\bfp'-\bfk)^2/2M +i\epsilon},
  \label{triangle2}
\end{eqnarray}
which is similar to Eq.~(\ref{triangle1}), but there is a crucial
difference. No nucleon pole appears in the lower half plane. Thus the
contribution can be evaluated by the residue of the pion pole in the
lower half plane and gives $\sim -\Lambda d\Lambda/4\pi^2$. This is
suppressed by a factor of $\Lambda/M$ compared with the diagram (b) in
Fig.~\ref{fournucleon}.


It is a general feature that the two-nucleon reducible (2NR) diagram
acquires the so-called ``IR enhancement,'' first noted by
Weinberg~\cite{Weinberg:1991um}, where the residue of the nucleon pole
gets an enhancement factor $M/\Lambda$. Note that the IR enhancement is
a consequence of the nonrelativistic kinematics. In addition, with the
residue of the nucleon pole, the pion propagator becomes
\begin{equation}
 \frac{i}
  {\left[-p^0+(\bfp+\bfk)^2/2M\right]^2-\omega_{k}^2}\; .
\end{equation}
Noting $p^0-\bfp^2/2M \ll \Lambda^2/M \ll \Lambda$, it is approximated by

\begin{equation}
 \sim \frac{-i}{\bfk^2+m_{\pi}^2}
\end{equation}
for $k\sim \Lambda$. That is, the effects of the pion can be represented
as the instantaneous potential.


Although we have shown that the mechanism of ``IR enhancement'' works in
the case of $m_{\pi} < \Lambda$, it is actually independent of this
assumption, and it also works in the case of $\Lambda < m_{\pi}$.

\subsubsection{Vertex corrections and the nucleon self-energy}

\begin{figure}
 \includegraphics[width=0.9\linewidth,clip]{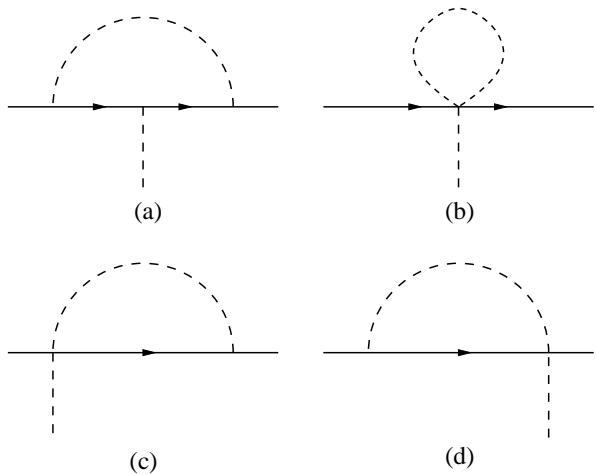}
 \caption{\label{vertex}Contributions to the $g_A$-vertex.}
\end{figure}
\begin{figure}
 \includegraphics[width=0.4\linewidth,clip]{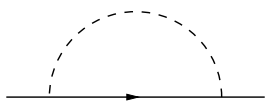}
 \caption{\label{selfenergy}Contributions to the self-energy of nucleon.}
\end{figure}

We also need to consider the renormalization of the $g_{A}$-term. The
relevant diagrams are depicted in Fig.~\ref{vertex}.  There are
contributions to the self-energy of the nucleon, shown in
Fig.~\ref{selfenergy}. These are potentially important to the
renormalization of the four-nucleon operators.



The examples in the previous subsection imply that the nucleon
self-energy diagram and the contributions to the nucleon-pion vertex do
not have IR enhancement. We will see shortly that these contributions
can be neglected in our leading order calculations. The key point is
that the appropriate dimensionless coupling constant for the
nucleon-pion coupling is, as we will explain in the next section
(Sec.~\ref{sec:spin-singlet}), given by $\gamma$,
\begin{equation}
 \gamma \equiv \frac{M\Lambda}{2\pi^2}\left(\frac{g_A}{2f}\right)^2.
  \label{gamma}
\end{equation}
The contributions from the nucleon-pion vertex to $\gamma$ should be
written in terms of $\gamma$.


There are four diagrams contributing to the nucleon-pion vertex shown in
Sec.~\ref{sec:order}.  The tadpole diagram (b) can be absorbed in the
redefinition of the coupling constant $g_{A}$. Each of the diagrams (c)
and (d) has an extra $1/M$ factor because the two-pion vertex comes from
the kinetic term $\bfD^2/2M$ so that they are suppressed compared to the
diagram (a).  The diagram (a) gives rise to the second term of the
right-hand side of the following RGE expressed in terms of $\gamma$,
\begin{equation}
 \frac{d\gamma}{dt} = 
  -\gamma -3 \left(\frac{\Lambda}{M}\right) \gamma^2.
  \label{alphacube}
\end{equation}
(The contributions from the nucleon wavefunction
renormalization is not considered here.) Note that the last term has an
explicit $\Lambda$-dependence.  In Eq.~(\ref{alpha}) the last term has
been neglected because $\Lambda/M$ is much smaller than one.


The self-energy diagram $\Sigma(p)$ shown in Fig~\ref{selfenergy} gives
contributions
\begin{equation}
 \Lambda\frac{d\Sigma(p)}{d\Lambda} = A + B p^0 + C \bfp^2
  +\mathcal{O}(\Lambda^2/M^2),
\end{equation}
where $A$, $B$, and $C$ are the constants that depend on the couplings
and the cutoff $\Lambda$. The constants $A$ and $C$ are canceled by the
counterterms, leaving the pole of the nucleon propagator intact. After
doing so, the propagator with this shell-mode
contribution becomes
\begin{equation}
 \left(1-B\frac{d\Lambda}{\Lambda}\right)^{-1}
  \frac{i}{p^0-\frac{\bfp^2}{2M}+i\epsilon},
\end{equation}
so that the contributions to the wave function renormalization constant
for the nucleon field, $Z_{N}$, can be written as
\begin{equation}
 \frac{dZ_{N}}{dt} = B.
\end{equation}
The order of magnitude estimate of $B$ gives
\begin{equation}
 B\sim \left(\frac{\Lambda}{M}\right)\gamma,
\end{equation}
so that the inclusion of the effect of wavefunction renormalization does
not alter the results of the leading order calculations.

\subsection{Averaged action with L-OPE}
\label{sec:lope}


We have seen that the leading contributions to the RGEs consist of the
one-loop 2NR diagrams with contact interactions and/or the instantaneous
pion exchanges. Actually these contributions can be generated by a much
simpler action than that in Eq.~(\ref{lagNR}). Thus we start with the
following ansatz for the averaged action $\Gamma_{\Lambda}$,
 \begin{eqnarray}
  \Gamma_{\Lambda}
   &=& 
   \Gamma^{(\Slash{\pi})}_{\Lambda}
   +
   \int d^4x
   \left\{
    -D_2^{(c)} m_{\pi}^2
    \mathcal{O}_{0}^{(c)}
   \right\}
   \nonumber \\
  &&+\frac{g_A^2}{4f^2} \int dt \int d^3x\; d^3y
   \bigg[
    \mathcal{O}^{(S)}(x,y)\nabla_x^2
    \nonumber \\
  &&+\mathcal{O}^{(T)}_{ii}(x,y)\nabla_x^2
   \nonumber \\
  &&-6\mathcal{O}^{(T)}_{ij}(x,y)
    \left(\der^x_i\der^x_j-\frac{1}{3}\delta_{ij}\nabla_x^2\right)
   \bigg]
   Y(\left|\bfx-\bfy\right|),
   \nonumber \\
  \label{lagNRreduced} 
 \end{eqnarray}
where we have introduced
\begin{eqnarray}
  \mathcal{O}^{(S)}(x,y)\!&=&\!
   \left(N^T(x)P_a^{(S)}N(y)\right)^\dagger\!\!\!
   \left(N^T(y)P_a^{(S)}N(x)\right),
   \nonumber \\
 &&\\
 \mathcal{O}^{(T)}_{ij}(x,y)\!&=&\!
   \left(N^T(x)P_i^{(T)}N(y)\right)^\dagger\!\!\!
   \left(N^T(y)P_j^{(T)}N(x)\right),
   \nonumber \\
\end{eqnarray}
and 
\begin{equation}
 Y(r) = \frac{1}{4\pi}\frac{e^{-m_\pi r}}{r}.
\end{equation}
Note that all the derivatives are now the usual ones, not the covariant
derivatives. It means that chiral symmetry is broken explicitly, but the
breaking is of higher order in $\Lambda/M$, as is seen from the
derivation. Note also that the operator corresponds to $D_2^{(c)}$ is
the same as that to $C_0^{(c)}$, but the former is a part of the
operator that emits/absorbs pions and is of higher order in the
$p$-expansion than the latter.


The effects of pions are represented as one-pion exchange
interactions. It is the L-OPE because the averaged action is defined
with the cutoff $\Lambda$, so that the S-OPE (with the momenta larger
than $\Lambda$) is included in the contact interactions.


The last term in Eq.~(\ref{lagNRreduced}) represents the tensor force of
L-OPE in the spin-triplet channel.

\section{Wilsonian RGEs for pionful NEFT $\mathcal{O}(p^2)$ in the
 $^1S_0$ and $^3S_1$-$^3D_1$ channels}
\label{sec:pionful}


We are now ready to calculate the RGEs for the two-nucleon system in the
S-waves. The formula Eq.~(\ref{sharpRGE}) can be used with slight
modification.
\begin{itemize}
 \item We introduce the dimensionless coupling constants $\gamma$
       defined in Eq.~(\ref{gamma}) for
       L-OPE, and $u$ and $u'$ for $D_2^{(c)}$, defined by
 \begin{equation}
  u\equiv \frac{M\Lambda^3}{2\pi^2}D_2^{(S)},
  \
  u'\equiv \frac{M\Lambda^3}{2\pi^2}D_2^{(T)}.
 \end{equation}
       Note that $g_A$ and $f$ appear in Eq.~(\ref{lagNRreduced}) only
       through the combination of $(g_{A}/f)^2$.
 \item The momentum-dependent factors associated with $\gamma$, $u$ and
       $u'$ are given by
       \begin{eqnarray}
F_{\gamma S}(p_{f},p_{i}) 
  \!&=&\!
  \left(\frac{-2\pi^2}{M\Lambda}\right)
  \half
  \left[
   \frac{r_{13}}{r_{13}\!+\!m_{\pi}^2}
   \!+\!\frac{r_{14}}{r_{14}\!+\!m_{\pi}^2}
  \right], \nonumber \\
&& \\
 F^{ij} _{\gamma T}(p_{f},p_{i})
  \!&=&\!
 \left(\frac{-6\pi^2}{M\Lambda}\right)
 \half
 \bigg[
  \frac{
   \delta^{ij}r_{13}- 2 p_{13}^i p_{13}^j
   }{r_{13}+m_{\pi}^2} 
   \nonumber \\
 &&\qquad\qquad\quad
  {}+
  \frac{
   \delta^{ij}r_{14}- 2 p_{14}^i p_{14}^j
   }{r_{14}+m_{\pi}^2} 
 \bigg],
 \\
F_u &=& F_{u'}=\frac{-2\pi^2 m_{\pi}^2}{M\Lambda^3}.
       \end{eqnarray}
 \item Since the $F_{\gamma S}$ and $F^{ij}_{\gamma T}$ have a bit more
       complicated momentum dependence than those in the pionless
       theory, the formula contains nontrivial integrations over angular
       variables. The part $F_A(p_i,\Lambda) F_B(\Lambda,p_f)$ in
       Eq.~(\ref{sharpRGE}) is replaced by $\langle
       F_A(p_i,\Lambda)F_B(\Lambda,p_f)\rangle$, where $\langle \cdots
       \rangle$ is defined as
       \begin{equation}
	\langle \cdots \rangle 
	 =\frac{1}{4\pi} \int d\Omega_{\hat{\bfk}}(\cdots),
       \end{equation}
       where $\Omega_{\hat{\bfk}}$ stands for the angular variables of
       $\bfk$.  See the original derivation of the formula for the
       details~\cite{Harada:2007ua}.
\end{itemize}

In the following, we assume that $m_{\pi} < \Lambda$ and expand in
powers of $m_{\pi}/\Lambda$. The case of $\Lambda < m_{\pi}$ is
discussed in Appendix~\ref{sec:lowlambda}.

\subsection{Spin-singlet channel}
\label{sec:spin-singlet}


In the spin-singlet channel, we have the following RGEs:
\begin{subequations}
\label{RGEsinglet}
\begin{eqnarray}
 \frac{dx}{dt}&=&
  -x
  -\Biggl[
  x^2+2xy+y^2+2xz+2yz+z^2
  \Biggr] \nonumber \\
 &&-2(x+y+z)\gamma -\gamma^2,
  \label{sing-x}
  \\
 \frac{dy}{dt}&=&
  -3y
  -\Biggl[
  \frac{1}{2}x^2+2xy+\frac{3}{2}y^2+yz-\frac{1}{2}z^2
  \Biggr] \nonumber \\
 &&-(x+2y)\gamma -\half \gamma^2, 
  \label{sing-y}
  \\
 \frac{dz}{dt}&=&
  -3z
  +\Biggl[
  \frac{1}{2}x^2+xy+\half y^2-xz-yz-\frac{3}{2}z^2
  \Biggr] \nonumber \\
 &&+(x+y-z)\gamma +\half \gamma^2,
  \label{sing-z}
  \\
 \frac{du}{dt}&=&-3u-2(x+y+z)(u-\gamma) -2u\gamma +2\gamma^2, 
\end{eqnarray}
\end{subequations}
and for $\gamma$,
\begin{equation}
 \frac{d\gamma}{dt}= -\gamma.\label{alpha}
\end{equation}
The first lines of Eqs~(\ref{sing-x}) -- (\ref{sing-z}) are the same as
those in the pionless calculations obtained in
Ref.~\cite{Harada:2006cw}. The terms in the second lines express how the
L-OPE is rearranged into the S-OPE when the floating cutoff is lowered.


We emphasize the choice made here of the dimensionless coupling constant
$\gamma$. There are several ways to make a dimensionless quantity from
the combination $(g_A/f)^2$, $M$, and $\Lambda$. Our choice is the one
for which the RGEs for the coupling constants of the contact
interactions do not have explicit $\Lambda$-dependence. If the explicit
$\Lambda$-dependence were present in the RGEs, the iterative property
(self-similarity) would be lost and the concept of fixed points would
become obscure. Since the unnaturally large scattering lengths in the
S-wave scattering are believed to be related to the nontrivial fixed
points, one needs to use such dimensionless variables that allow fixed
points.


The nontrivial fixed point of the RGEs~(\ref{RGEsinglet}) and
Eq.~(\ref{alpha}) relevant to
the real two-nucleon system is found to be
\begin{equation}
(x^\star, y^\star, z^\star, u^\star, \gamma^\star) 
 = \left(-1, -\half, \half, 0, 0\right),
\end{equation}
which is identified with that found in the pionless NEFT, given in
Eq.~(\ref{ntfpsinglet}). 


We linearize the RGEs around the fixed point, and find the following
eigenvalues (scaling dimensions) and the corresponding eigenvectors:
\begin{eqnarray}
 \nu_1&=&+1:\ 
 u_1=\left(
      \begin{array}{c}
       1\\
       1\\
       -1\\
       0\\
       0
      \end{array}
      \right),\quad
 \nu_2=-1:\ 
 u_2=\left(
      \begin{array}{c}
       0\\
       -1\\
       1\\
       0\\
       0
      \end{array}
     \right),
 \nonumber \\
 \nu_3&=&-2:\ 
 u_3=\left(
      \begin{array}{c}
       2\\
       -1\\
       -2\\
       0\\
       0
      \end{array}
     \right), \quad
 \nu_4=-1:\ 
 u_4=\left(
      \begin{array}{c}
       0\\
       0\\
       0\\
       1\\
       0
      \end{array}
     \right).
 \nonumber \\
\end{eqnarray}
The eigenvalues $\nu_1$, $\nu_2$, and $\nu_3$ and corresponding
eigenvectors $u_1$, $u_2$, and $u_3$ can be identified with those found
in the pionless theory. Therefore, the power counting is not modified
by the inclusion of the pions; only the one relevant operator ($u_1$)
should be resummed to all orders.


Note that the eigenvalue problem is five dimensional and one thus
expects five pairs of eigenvalues and eigenvectors. It is easily seen
that the eigenvalue $-1$ is triply degenerate. Two of the (nonzero)
eigenvectors are $u_2$ and $u_4$, but the third one does not exist. This
is not a mathematical inconsistency, however. Although we do not
understand very well the reason why the third eigenvector is missing, it
clearly has something to do with the $\gamma$-direction, because the
vector in the $\gamma$-direction cannot be expressed as a linear
combination of $u_{i}\mbox{'s}\  (i=1,\cdots, 4)$.


Eq.~(\ref{alpha}) shows that the L-OPE is irrelevant. It implies that
the L-OPE should be treated as a perturbation. Note that there is a
typical scale in the pionful NEFT,
$\Lambda_{NN}$,
\begin{equation}
 \Lambda_{NN} = \frac{4\pi}{M}\left(\frac{2f}{g_A}\right)^2,
  \label{lambdaNN}
\end{equation}
and our $\gamma$ is related to it as
\begin{equation}
 \gamma(\Lambda) = \frac{2}{\pi}
  \left(\frac{\Lambda}{\Lambda_{NN}}\right).
\end{equation}
Kaplan, Savage, and Wise~\cite{Kaplan:1998tg, Kaplan:1998we} regard
$p/\Lambda_{NN}$ as an expansion parameter. Our finding is consistent
with their approach.

\subsection{Spin-triplet channel}


In the spin-triplet channel, we have the following RGEs:
\begin{widetext}
\begin{subequations}
\label{RGEtriplet}
\begin{eqnarray}
 \frac{dx'}{dt}&=&
  -x'
  -\Biggl[
  {x'}^{2}+2x'y'+{y'}^{2}+2x'z'+2y'z'+{z'}^{2}+2{w'}^{2}
  \Biggr]\nonumber \\
 &&-2(x'+y'+z'-4w')\gamma -9\gamma^2,
  \label{trip-x}\\
 \frac{dy'}{dt}&=&
 -3y'
 -\Biggl[
 \frac{1}{2}{x'}^{2}+2x'y'+\frac{3}{2}{y'}^{2}+y'z'-\frac{1}{2}{z'}^{2}
 +{w'}^{2}
 \Biggr]\nonumber \\
 &&-(x'+2y')\gamma +\frac{7}{2}\gamma^2, 
  \label{trip-y}\\
 \frac{dz'}{dt}&=&
 -3z'
 +\Biggl[
 \frac{1}{2}{x'}^{2}+x'y'+\half {y'}^{2}-x'z'-y'z'-\frac{3}{2}{z'}^{2}
 +{w'}^{2}
 \Biggr]\nonumber \\
 &&+(x'+y'-z'-4w')\gamma + \frac{9}{2}\gamma^2, 
  \label{trip-z}\\
 \frac{dw'}{dt}&=&
 -3w'
 -\Biggl[
 x'w'+y'w'+z'w'
 \Biggr]\nonumber \\
 &&+\frac{1}{5}(2x'+2y'+2z'-9w')\gamma +2\gamma^2, 
  \label{trip-w}\\
 \frac{du'}{dt}&=&
 -3u'-2(x'+y'+z')u' +2(x'+y'+z'-4w')\gamma -2u'\gamma +18\gamma^2.
 \label{trip-u}
\end{eqnarray} 
\end{subequations} 
\end{widetext}
Here, again, the first lines of Eqs~(\ref{trip-x}) -- (\ref{trip-w}) are
the same as those in the pionless calculations, obtained in
Ref.~\cite{Harada:2006cw}, while the second lines are the contributions
from L-OPE. 


Note that the magnitude of the coefficients of $\gamma^2$ is large
compared to those for the spin-singlet channel. This is the effect of
the tensor part of L-OPE.


The nontrivial fixed point of the RGEs~(\ref{RGEtriplet}) and
Eq.~(\ref{alpha}) relevant to
the real two-nucleon system is found to be
\begin{equation}
(x^\star, y^\star, z^\star, w^\star, u^\star, \gamma^\star) 
 = \left(-1, -\half, \half, 0, 0, 0\right),
\end{equation}
which is the same as that found in the pionless NEFT, given in
Eq.~(\ref{ntfptriplet}). 


The RGEs linearized around the fixed point lead to the following set of
eigenvalues and the eigenvectors:
\begin{eqnarray}
 \nu_1&=&+1:\ 
  u_1=\left(
       \begin{array}{c}
	1\\
	1\\
	-1\\
	0\\
	0\\
	0
       \end{array}
      \right),\quad
  \nu_2=-1:\ 
  u_2=\left(
       \begin{array}{c}
	0\\
	-1\\
	1\\
	0\\
	0\\
	0
       \end{array}
      \right),
  \nonumber \\
  \nu_3&=&-2:\ 
  u_3=\left(
       \begin{array}{c}
	2\\
	-1\\
	-2\\
	0\\
	0\\
	0
       \end{array}
      \right),\quad
 \nu_4=-2:\ 
  u_4=\left(
       \begin{array}{c}
	0\\
	0\\
	0\\
	1\\
	0\\
	0
       \end{array}
      \right),
  \nonumber \\
 \nu_5&=&-1:\ 
  u_5=\left(
       \begin{array}{c}
	0\\
	0\\
	0\\
	0\\
	1\\
	0
       \end{array}
      \right)
\end{eqnarray}
Here, again, the eigenvalues $\nu_1$, $\nu_2$, $\nu_3$ and $\nu_4$ and
corresponding eigenvectors $u_1$, $u_2$, $u_3$ and $u_4$ can be
identified with those found in the pionless theory. Therefore, in spite
of the presence of the tensor force in this channel, the power counting
is not modified.  As in the spin-singlet channel, only the one relevant
($u_1$) should be resummed to all orders.


As in the spin-singlet channel, the eigenvalue $-1$ is triply
degenerate, but one eigenvector is missing.

\subsection{The effects of pions}

In this section, we discuss the effects of pions to the RGEs and thus to
the power counting.


As we discussed in Sec.~\ref{sec:pionless}, the scaling dimensions at
the nontrivial fixed point determine the power counting of NEFT in the
S-waves in the two-nucleon sector. In the previous sections, we have
seen that the nontrivial fixed points as well as the scaling dimensions
at them in both $^1S_0$ and $^3S_1$--$^3D_1$ channels remain the same
even when pions are included as explicit degrees of freedom. Therefore
the power counting for the contact operators in the pionful theory is
the same as that of the pionless theory.


The L-OPE does not affect the location of the fixed points and the
scaling dimensions there, but the details of the flows. The ``strong''
tensor force in the spin-triplet channel does not change the properties
of the nontrivial fixed point, but gives rise to the strong
$\gamma$-dependence (because of the large factor ``6'' in
Eq.~(\ref{lagNRreduced})) in the channel and modifies the flow
considerably compared to the spin-singlet channel.


The effects can be read off from the $\gamma$-dependence of the RGEs. To
see the effects in more details, let us introduce new variables,
\begin{equation}
 X=x+y+z,\quad Y=y+z,\quad Z=y-z,
\end{equation}
and rewrite the RGEs in Eq.~(\ref{RGEsinglet}) for the spin-singlet
channel as follows:
\begin{subequations}
\begin{eqnarray}
 \frac{dX}{dt}&=& -X\! -2Y\! -XY -X^2\! -(2X+Y)\gamma - \gamma^2, \\
 \frac{dY}{dt}&=& -(3+X+\gamma)Y, \label{y} \\
 \frac{dZ}{dt}&=& -3Z -X(X-Y+2Z) -(2X-Y+2Z)\gamma
  - \gamma^2, \nonumber \\
 && \\
 \frac{du}{dt}&=& -3u-2Xu -2(X+u)\gamma +2\gamma^2.
\end{eqnarray}
\end{subequations}
We see that the RGEs for $(X,Y, \gamma)$ form a closed subset and can be
solved without knowing the flow for $Z$ and $u$. In these variables, the
nontrivial fixed point is given by $(X^\star, Y^\star, \gamma^\star) =
(-1,0,0)$. 


Similarly, we introduce 
\begin{equation}
 X'=x'+y'+z',\quad Y'=y'+z',\quad Z'=y'-z',
\end{equation}
in the spin-triplet channel, and the RGEs can be rewritten as
\begin{subequations}
\begin{eqnarray}
 \frac{dX'}{dt}&=& -X' -2Y' -X'Y' -{X'}^2 -2{w'}^2
  \nonumber \\
 &&-(2X'+Y'-4w')\gamma - \gamma^2, \\
 \frac{dY'}{dt}&=& -(3+X')Y' -(Y'+4w')\gamma
  +8\gamma^2,  
  \label{yprime}\\
 \frac{dZ'}{dt}&=& -3Z' -X'(X'-Y'+2Z') -2{w'}^2
  \nonumber \\
 &&-(2X'-Y'+2Z'-4w')\gamma - \gamma^2, \\
 \frac{dw'}{dt}&=& -3w' -X'w' +\frac{1}{5}(2X'-9w')\gamma 
  \nonumber \\
 &&+2\gamma^2, \\
 \frac{du'}{dt}&=& -3u'-2Xu' +2(X'-u'-4w')\gamma 
  \nonumber \\
 &&+18\gamma^2. 
\end{eqnarray} 
\end{subequations}
The RGEs for $(X',Y',w', \gamma)$ form a closed subset, and the
nontrivial fixed point is given by $({X'}^\star, {Y'}^\star,
{w'}^\star,\gamma^\star) = (-1,0,0,0)$.

\begin{figure}
 \includegraphics[width=0.9\linewidth,clip]{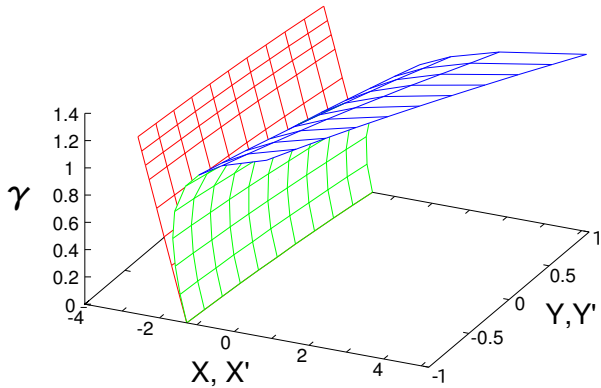}
 \caption{\label{surfaces} (Color online) The surfaces which separate
 the initial values into those from which the flows go to the strong
 coupling phase (the left side), and those to the weak coupling phase
 (the right side). The surface on the right is for the spin-triplet
 channel, and the other on the left for the spin-singlet channel. All
 the points have the initial values with $w'=0$ for the spin-triplet
 channel.}
\end{figure}

Thanks to the similarity of the RGEs in both channels, we can project
the flows onto the $(X,Y,\gamma)$ and $(X', Y', \gamma)$ space and
compare them in the single plot. In Fig.~\ref{surfaces}, we have drawn
the surfaces which separate the initial values into two regions: the
region from which the flows go to the strong coupling phase, and the
other to the weak coupling phase. It is evident that the region of
initial values to the strong coupling phase is larger in the
spin-triplet channel than in the spin-singlet channel. 

\begin{figure}[h]
 \includegraphics[width=0.9\linewidth,clip]{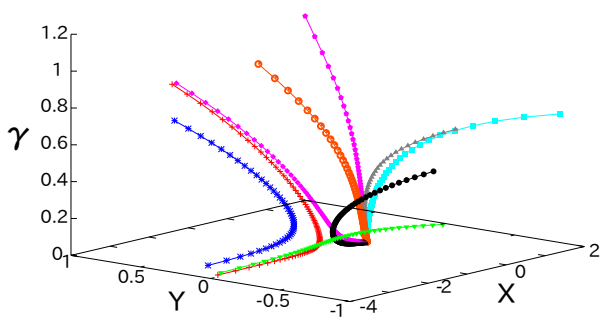}
 \caption{\label{singflow} (Color online) Typical RG flow lines in the
 spin-singlet channel. The intervals between points indicate how fast
 they flow.}
\end{figure}

\begin{figure}[h]
 \includegraphics[width=0.9\linewidth,clip]{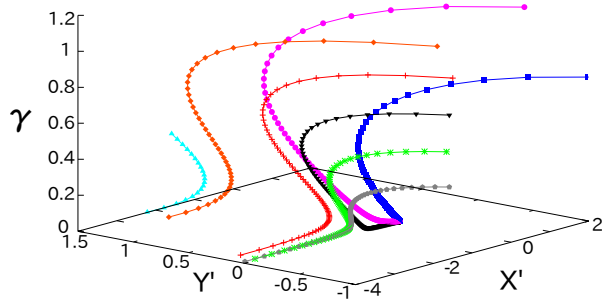}
 \caption{\label{tripflow} (Color online) Typical RG flow lines in the
 spin-triplet channel with the initial value of $w'$ set equal to
 zero. The intervals between points indicate how fast they flow. Note
 that the strong ``dragging force'' in the $Y'$ direction.}
\end{figure}

In Figs.~\ref{singflow} and \ref{tripflow}, we show some typical flows.
Note that the large coefficient of the $\gamma^2$ term of
Eq.~(\ref{yprime}), in comparison with Eq.~(\ref{y}) results in the
large bending of the flow lines in the $Y'$ direction.

\section{Summary and discussions}
\label{sec:summary}

\subsection{Summary}


In this paper, we consider the pionful NEFT in the two-nucleon sector in
the S-waves in the leading order of the nonrelativistic approximation in
order to study the power counting issue from the Wilsonian RG point of
view. We show that the leading order contributions to the RGEs come from
the two-nucleon-reducible diagrams, and the pion propagators are dominated
by the instantaneous Yukawa potential.


The separation of the pion exchange contributions into
the L-OPE and the S-OPE is emphasized on the basis of the general
effective field theory philosophy. The L-OPE is expressed as the Yukawa
potential in the averaged action, while the S-OPE is included in the
contact interactions along with the other short-distance effects.


We derive the RGEs for the spin-singlet and spin-triplet channels from
the effective averaged action ansatz up to including the
$\mathcal{O}(p^2)$ in the expansion of momenta and the pion mass. The
nontrivial fixed point of physical importance is found to be the same as
that in the pionless NEFT in each channel. The eigenvalues (scaling
dimensions) and the corresponding eigenoperator of the linearized RGEs
around the fixed point are also shown to be the same.  That is, there is
one relevant contact operator to be resummed. The other operators should
be treated as perturbations. The L-OPE is also treated as a
perturbation. A part of the S-OPE contained in the relevant operator is
resummed to all orders.


We emphasize that the effects of pions do not alter the scaling
dimensions and hence the power counting. The pions affect the details of
the RG flows. We show that in the spin-triplet channel the flow is more
affected by the pions and the region of the initial values that flow to
the strong coupling phase is larger than in the spin-singlet channel.

We believe that the difference between these channels eventually leads
to the existence of the bound state (the deuteron) in the spin-triplet
channel, and the nonexistence in the spin-singlet channel.

\subsection{Comments on the related works}

In the following, we discuss the relation of the present paper to the
relevant works in the literature.


As stated in Introduction, our work is very closely related to the work
by Beane, Kaplan, and Vuorinen~\cite{Beane:2008bt}, who, working with
the PDS renormalization scheme, introduce a regularization mass scale
$\lambda$ to separate the pion exchange into its long-distance and
short-distance parts, and the short-distance part is represented as
contact interactions. The separation of the pion exchange into two parts
is essential to improve the convergence of the EFT expansion in the
spin-triplet channel and is similar to our Wilsonian RG analysis, with
the regularization scale $\lambda$ corresponding to our floating cutoff
$\Lambda$. 

There are however several points to be addressed: (i) Even though they
regard $\lambda$ as a low-energy scale, they actually consider
high-momentum values in their numerical calculation, ranging from 600 to
1000 MeV.  (ii) They employ the PDS renormalization scheme simply
assuming the modification does not affect the power counting.  (iii) The
renormalization scale $\mu$ and the regularization scale $\lambda$ seem
to play a similar role in reordering the EFT expansion but they are
treated independently. As a result, $\lambda$ becomes just a new
parameter. (iv) They consider the regularization only for the
spin-triplet channel because of the singular $1/r^3$ potential, but not
for the spin-singlet channel.

From our point of view, these may be seen as follows: (i) Our separation
scale $\Lambda$ is smaller than the physical cutoff $\Lambda_0 \approx
400$ MeV, so that it can be consistently regarded as a low-energy
scale. (ii) We show in this paper that the nontrivial fixed points as
well as the scaling dimensions are the same as those in the pionless
theory. We have shown that the scaling dimensions near the nontrivial
fixed points in the pionless theory correspond to those of the PDS
ones~\cite{Harada:2007ua}. (iii) In our Wilsonian RG analysis, there is
only one scale $\Lambda$, which plays the role of $\mu$, renormalization
scale, in the RGEs, and simultaneously the role of $\lambda$, the
separation scale. (iv) We think that one should consider such a
regularization for both channels from a general EFT point of view.


The effects of pions in the Wilsonian RG context have been studied by
Birse~\cite{Birse:2005um, Birse:2007sx}. He shows that the range of
momenta in which the expansion converges is very small in the
spin-triplet channel and claims that the pion exchanges should be
resummed. He then employs the distorted-wave RG~\cite{Barford:2002je}
and finds that the scaling dimensions shift by $-1/2$ (in our
convention) in the spin-triplet channel at the nontrivial fixed
point. Even at the trivial fixed point, the scaling dimensions are found
to shift from the canonical values. He also notes that the scale
$\lambda_{\pi}$ (which is the same as $\Lambda_{NN}$ in
Eq.~(\ref{lambdaNN})) is numerically small ($\sim 2m_{\pi}$) so that the
resummation is justified.

We note that he does not introduce the separation scale $\Lambda$ for
the pion exchange consistently with the contact interactions, but
introduces an additional regularization scale ($R$) and keeps it fixed
when studying the RG flows. Thus his distorted-wave RGE does not take
into account the contributions from the OPE to the contact interactions
(diagrams (b) to (d) in Fig.~\ref{fournucleon}) at all. As a result, his
RGE does not have a smooth transition to that of the pionless theory,
where all the pion exchange effects are represented as contact
interactions. That is, the pion exchanges never decouple. In contrast,
our RGEs have a smooth transition to those of the pionless theory, as
shown in Appendix~\ref{sec:lowlambda}, and the pions decouple as they
should. In addition, because of the iterative property of the RGEs, the
definition of the dimensionless coupling, $\gamma$, is uniquely
determined, and it leads to the perturbative treatment of L-OPE, as
explained at the end of Sec.~\ref{sec:spin-singlet}. 


Fleming et al.~\cite{Fleming:1999ee} also report the nonconvergence of
the EFT expansion in the P-waves (the $^3P_{0}$ channel). Although we
have not explicitly calculated the RGEs in this channel, we think that a
similar machinery also works there. The difference between the S-waves
and the P-waves is that the physical NN system is near the nontrivial
fixed point in the S-waves, while in the P-waves it seems to be near the
trivial fixed point. Although the scaling dimensions obtained in the PDS
scheme with the pole at $D=3$ subtracted are shown to be the same as
those at the nontrivial fixed point in the S-waves, they correspond
neither to those at the nontrivial fixed point nor to those at the
trivial fixed point in the P-waves~\cite{Harada:2007ua}. Thus the simple
PDS with the pole at $D=3$ subtracted should not be used.

\subsection{Prospects of future research}


Finally we would like to make a comment on a possible implementation of
the findings of the present paper into a more tractable way of
calculating the physical amplitudes to higher orders. The Wilsonian RG
method with the momentum cutoff is theoretically transparent but
practically too complicated to do higher order calculations. A simple
but powerful scheme that is also consistent with our results is
desired. Such a scheme would employ the dimensional
regularization. Since dimensional regularization does not have a natural
separation scale in itself, one should introduce it by hand. Thus it
would be very similar to the BKV prescription. It seems necessary,
however, to make a connection between the renormalization scale $\mu$
and the separation scale $\lambda$ so that we have a consistent RGEs
with those obtained in the present paper. Work in this direction is now
in progress.

\appendix
\section{RGEs for the case of $\Lambda < m_{\pi}$}
\label{sec:lowlambda}

In Sec.~\ref{sec:pionful}, we have derived the RGEs for the case
$m_{\pi}< \Lambda$ by expanding the contributions in powers of
$m_{\pi}/\Lambda$. But the pionful NEFT is valid also for the case
$\Lambda < m_{\pi}$. In this Appendix we present the RGEs for the case
$\Lambda < m_{\pi}$. The diagrams which contribute to the RGEs are the
same. The difference is that the contributions are now expanded in
powers of $\Lambda/m_{\pi}$.

In the spin-singlet channel, we have
\begin{subequations}
\label{RGEsinglet_less}
\begin{eqnarray}
 \frac{dx}{dt}&=&
\mbox{(first line of Eq.~(\ref{sing-x}))}
\nonumber \\
 &&-2(x+y+z+ \tilde{u})\tilde{\gamma} -\tilde{\gamma}^2,
  \\
 \frac{dy}{dt}&=&
 \mbox{(first line of Eq.~(\ref{sing-y}))}
\nonumber \\
 &&-(2x+3y+z+2\tilde{u})\tilde{\gamma} -\frac{3}{2} \tilde{\gamma}^2, 
  \\
 \frac{dz}{dt}&=&
 \mbox{(first line of Eq.~(\ref{sing-z}))}
\nonumber \\
 &&+(x+y-z+\tilde{u})\tilde{\gamma} +\half \tilde{\gamma}^2,
  \\
 \frac{d\tilde{u}}{dt}&=&
 -\tilde{u}-2(x+y+z)\tilde{u} 
 -2\tilde{u}\tilde{\gamma},
\end{eqnarray}
\end{subequations}
where we have introduced new notations,
\begin{equation}
 \tilde{\gamma}=\frac{\Lambda^2}{m_{\pi}^2}\gamma,
  \quad
  \tilde{u} = \frac{m_{\pi}^2}{\Lambda^2}u.
\end{equation}
Now the RGE for $\tilde{\gamma}$ is given by
\begin{equation}
 \frac{d\tilde{\gamma}}{dt}=-3\tilde{\gamma}.
  \label{alphatilde}
\end{equation}

Similarly, in the spin-triplet channel, we have
\begin{subequations}
\label{RGEtriplet_less}
\begin{eqnarray}
 \frac{dx'}{dt}&=&
  \mbox{(first line of Eq.~(\ref{trip-x}))}
  \nonumber \\
 &&-2(x'+\tilde{u}'+y'+z'-4w')\tilde{\gamma} -9\tilde{\gamma}^2,
  \\
 \frac{dy'}{dt}&=&
 \mbox{(first line of Eq.~(\ref{trip-y}))}
\nonumber \\
 &&-(2x'\!+2\tilde{u}'\!+3y'\!+z'\!-4w')\tilde{\gamma} 
  -\frac{11}{2}\tilde{\gamma}^2, 
  \\
 \frac{dz'}{dt}&=&
 \mbox{(first line of Eq.~(\ref{trip-z}))}
 \nonumber \\
 &&+(x'+\tilde{u}'+y'-z'-4w')\tilde{\gamma} + \frac{9}{2}\tilde{\gamma}^2, 
  \\
 \frac{dw'}{dt}&=&
 \mbox{(first line of Eq.~(\ref{trip-w}))}
 \nonumber \\
 &&+(2x'+2\tilde{u}'+2y'+2z'-w')\tilde{\gamma} +2\tilde{\gamma}^2, 
  \\
 \frac{d\tilde{u}'}{dt}&=&
 -\tilde{u}'-2(x'+y'+z')\tilde{u}' -2\tilde{u}'\tilde{\gamma}.
\end{eqnarray} 
\end{subequations} 


In the case of $\Lambda < m_{\pi}$, the operator corresponding
$D_2^{(c)}$ cannot be distinguished with the operator corresponding to
$C_0^{(c)}$, because these operators are of the same form to this order
and the pion mass is not a small parameter to be expanded. Thus they
appear only though the combinations $C_0^{(c)} + m^2_{\pi}D_2^{(c)}$,
or, $\chi\equiv x +\tilde{u}$ and $\chi'\equiv x'+ \tilde{u}'$. 
In terms
of these variables, the RGEs can be rewritten as
\begin{subequations}
\label{RGEsinglet_less2}
\begin{eqnarray}
 \frac{d\chi}{dt}&=&
  \mbox{(first line of Eq.~(\ref{sing-x}) with $x$ $\to$ $\chi$)}
  \nonumber \\
 &&-2(\chi+y+z)\tilde{\gamma} -\tilde{\gamma}^2,
  \\
 \frac{dy}{dt}&=&
 \mbox{(first line of Eq.~(\ref{sing-y}) with $x$ $\to$ $\chi$)}
 \nonumber \\
 &&-(2\chi+3y+z)\tilde{\gamma} -\frac{3}{2} \tilde{\gamma}^2, 
  \\
 \frac{dz}{dt}&=&
  \mbox{(first line of Eq.~(\ref{sing-z}) with $x$ $\to$ $\chi$)}
 \nonumber \\
 &&+(\chi+y-z)\tilde{\gamma} +\half \tilde{\gamma}^2,
\end{eqnarray}
\end{subequations}
for the spin-singlet channel, and 
\begin{subequations}
\label{RGEtriplet_less2}
\begin{eqnarray}
 \frac{d\chi'}{dt}&=&
   \mbox{(first line of Eq.~(\ref{trip-x}) with $x'$ $\to$ $\chi'$)}
  \nonumber \\
 &&-2(\chi'+y'+z'-4w')\tilde{\gamma} -9\tilde{\gamma}^2,
  \\
 \frac{dy'}{dt}&=&
 \mbox{(first line of Eq.~(\ref{trip-y}) with $x'$ $\to$ $\chi'$)}
 \nonumber \\
 &&-(2\chi'+3y'+z'-4w')\tilde{\gamma} 
  -\frac{11}{2}\tilde{\gamma}^2, 
  \\
 \frac{dz'}{dt}&=&
 \mbox{(first line of Eq.~(\ref{trip-z}) with $x'$ $\to$ $\chi'$)}
 \nonumber \\
 &&+(\chi'+y'-z'-4w')\tilde{\gamma} + \frac{9}{2}\tilde{\gamma}^2, 
  \\
 \frac{dw'}{dt}&=&
 \mbox{(first line of Eq.~(\ref{trip-w}) with $x'$ $\to$ $\chi'$)}
 \nonumber \\
 &&+(2\chi'+2y'+2z'-w')\tilde{\gamma} +2\tilde{\gamma}^2, 
\end{eqnarray} 
\end{subequations} 
for the spin-triplet channel. Here we have now included the other terms
that we neglected in the ansatz to the order $\mathcal{O}(p^2)$, such as
terms proportional to $m_{\pi}^4$. In a similar way, we may consider
that $\chi$ and $\chi'$ contain the terms proportional to $m_{\pi}^2$,
but also terms of all order in the expansion in $m_{\pi}^2$. These
$\chi$ and $\chi'$ should be compared to the couplings $x$ and $x'$ in
the pionless NEFT.


The new coupling $\tilde{\gamma}$ is a natural measure of the strength
of the pion exchange for $\Lambda < m_{\pi}$, as $\gamma$ is for
$m_{\pi}< \Lambda$.  Note that the RGE for the $\tilde{\gamma}$,
Eq.~(\ref{alphatilde}), shows that the L-OPE is more irrelevant and the
effects of pions on the contact interactions thus become negligible
very rapidly in this region. This implies that one may put
$\tilde{\gamma}= 0$ as a good approximation. In this way, the RGEs of
the pionful NEFT is smoothly connected to those of the pionless NEFT.

\section{The case of QED}
\label{sec:qed}

In this Appendix, we briefly discuss the case of QED in the
nonrelativistic region (NRQED). As a concrete example, we have a
hydrogen atom (or, electron-proton scattering ) in mind, and
we are interested in the low-energy region where even the electron
behaves as a nonrelativistic particle.


The system consists of a nonrelativistic proton and an electron
interacting by exchanging photons. The Lagrangian is similar to that of
NEFT with propagating pions, Eq.~(\ref{lagNR}). Note that we also
include contact interactions of protons and electrons. By a similar
analysis, one easily finds that the IR enhancement takes place for the
proton-electron reducible diagrams, giving the leading order
contributions to the RGEs. There,(the time-time component of) the photon
propagator is replaced with the instantaneous Coulomb potential. (It is
independent of the choice of the gauge.) Effectively, the RGEs are
generated by the averaged action consisting of the contact interactions
and the instantaneous Coulomb interaction, such as
\begin{eqnarray}
 &&-\frac{e^2}{2}\int\! dt\! \int\! d^3x\; d^3y
  (e^T(t,\bfx)\sigma_2p(t,\bfy))^\dagger
  (e^T(t,\bfx)\sigma_2p(t,\bfy))
  \nonumber \\
  &&\qquad \times 
   C\left(\left|\bfx-\bfy\right|\right),
\end{eqnarray}
where $C(r)$ is the Coulomb potential,
\begin{equation}
 C(r) =\frac{1}{4\pi r},
\end{equation}
similar to that of NEFT, Eq.~(\ref{lagNRreduced}). The Coulomb
interaction here should be considered as the long-distance part of the
Coulomb interaction (L-Coulomb), while the short-distance part
(S-Coulomb) is included in the contact interactions.


The difference between the pionful NEFT and the NRQED comes from the
following facts: (i) the photon is massless while the pion is massive,
(ii) the Coulomb interaction is not a derivative coupling, while
the NN$\pi$ is, and (iii) the electromagnetic coupling $e^2$ is
dimensionless while the NN$\pi$ coupling $(g_A/2f)^2$ is dimensionful.


For the NRQED RGEs, we introduce the dimensionless coupling constant
$\gamma_{QED}$,
\begin{equation}
 \gamma_{QED} \equiv e^2 \frac{M}{\Lambda},
\end{equation}
where $M$ is the reduced mass of the proton and the electron. This
particular $\Lambda$-dependence is required by the condition that the
RGEs for the contact interactions do not contain the explicit
$\Lambda$-dependence, as in Sec.~\ref{sec:spin-singlet} for the pionful
NEFT. The RGE for $\gamma_{QED}$ in the leading order is given by
\begin{equation}
 \frac{d\gamma_{QED}}{dt} = + \gamma_{QED}.
\end{equation}
(Note that the usual beta function contribution vanishes, $de^2/dt = 0$,
because we are in a region where $\Lambda$ is smaller than the electron
mass.) We see that the L-Coulomb interactions become more important in
the infrared, in strong contrast to the L-OPE. Photons do not decouple
as they should not. There is no ``photonless'' QED.


The contact interactions lead to the similar RGEs to those of the NEFT.
There is a nontrivial fixed point similar to the one found in the
NEFT. Since there seems to be no fine-tuning for NRQED, it is natural to
consider that the physical system is close to the trivial fixed
point. Therefore, the S-Coulomb is irrelevant and should be treated as a
perturbation.


In the pionful NEFT case, the L-OPE becomes rapidly irrelevant for
$\Lambda < m_{\pi}$, as is shown in Appendix~\ref{sec:lowlambda}. On the
other hand, NRQED does not have such a region where the L-Coulomb
becomes irrelevant because the photon is massless.


Therefore, we conclude that the L-Coulomb interaction is
nonperturbative, and the contact interactions including the S-Coulomb
interaction are perturbative. It is interesting to compare this with the
case of the pionful NEFT, where the L-OPE is perturbative while (a part
of) the S-OPE should be resummed to all orders.

\begin{acknowledgments}
 The authors thank M.~C.~Birse, O.~J.~Rosten, and Ulf-G. Mei{\ss}ner for
 the commnets and the suggestions. 

 This work was supported by JSPS KAKENHI (Grand-in-Aid for Scientific
 Research (C)) (22540286).
\end{acknowledgments}

\bibliography{NEFT,NPRG,NLS}

\end{document}